\documentclass[11pt]{article}

\usepackage[T1]{fontenc}
\usepackage[utf8]{inputenc}
\usepackage{lmodern}

\usepackage{amsmath,amssymb}
\usepackage{microtype}
\usepackage[hidelinks]{hyperref}

\title{Determinism and Indeterminism as Model Artefacts:\\
Toward a Model-Invariant Ontology of Physics}

\author{David Nolland}

\date{}

\begin{document}
\maketitle

\begin{abstract}

This paper argues that, where deterministic and indeterministic formulations agree on empirically corroborated structure while differing in their completion of structure that remains empirically underdetermined, features invariant across these formulations warrant ontological commitment, whereas any warrant attaching to features that vary across them is categorically more provisional. Because deterministic and indeterministic completions are always formally available, while additional grounds for preferring particular completions are contingent and defeasible, empirically corroborated structure alone cannot determine whether reality is fundamentally deterministic or indeterministic, or whether that very distinction belongs only to specific representations of it.

On this basis, I advance a model-invariance criterion according to which structural features are candidates for realist commitment insofar as they are empirically corroborated within a regime and preserved across the admissible formulations of that structure. I show how such invariant content can be characterised in mathematically intelligible ways, yielding a fallibilist form of structural realism grounded in model-invariance rather than in the specific form of any given mathematical representation. Features such as conservation relations, invariant measures, symmetries, and probability or correlation structures over observable histories may satisfy this criterion. By contrast, the localisation of modal selection---whether in fine-grained initial conditions, stochastic outcomes, or state-update and collapse mechanisms---may vary across empirically admissible formulations and therefore constitute a gauge-like representational freedom.

This framework offers a realist ontology for modern physics that combines empirical openness with resistance to metaphysical overreach.

\end{abstract}

\section{Introduction}

Since the Scientific Revolution, it has been common to treat determinism and indeterminism as marking a deep divide in our fundamental understanding of the natural world. In a deterministic formulation, the state at one time together with the dynamical law fixes a unique history (Butterfield 2005; Hoefer 2026); the information selecting that history can accordingly be represented in the state specification, often as an initial condition. In an indeterministic formulation, by contrast, trajectory-selection is represented through stochastic outcomes distributed throughout the history.

There is a persistent temptation to identify properties of our most fundamental representations with properties of reality itself, and hence to insist that one or other of these conceptions corresponds to the basic structure of reality: that whichever is instantiated by the most fundamental description available to us at a given time must therefore be ontological. Quantum mechanics thus reversed the dominant metaphysical default: where classical mechanics had lent itself to deterministic readings of nature, standard interpretations of quantum mechanics favoured indeterministic ones.

The situation is complicated when we realise that there are broad classes of cases, in both classical and quantum mechanics, in which deterministic and indeterministic descriptions are empirically equivalent, and in which the localisation of modal selection can be transposed between deterministic and indeterministic representations. The question then arises: even where the quest for theoretical unification gives us reason to favour one particular representation as more fundamental, under what circumstances does that preference warrant treating its deterministic or indeterministic character as ontological? Or, to take this even further, could the very distinction between determinism and indeterminism be a feature of our representations that need not correspond to a distinction in reality at all? And if this is the case, what features of our models are we justified in taking as ontological?

Work by Werndl (2009, 2011) and Ornstein \& Weiss (1991) has established observational equivalence between deterministic and indeterministic formulations for a wide range of dynamical systems under suitable coarse-grainings. The general picture is that a formally deterministic system can admit an observationally equivalent stochastic description when attention is restricted to the coarse-grained preparation--measurement correlation structure. For important classes of chaotic---and even some non-chaotic---systems, this equivalence can persist at every level of empirical resolution, in the appropriate sense of observational equivalence, and both models can be of a kind commonly used in science (Werndl 2009, 2011).

In the paradigmatic chaotic cases, this observational duality has an intuitive dynamical basis: arbitrarily fine differences in initial conditions can be amplified into macroscopically distinct outcomes, so that no finite increase in spatial resolution can prevent unresolved structure from eventually becoming dynamically consequential. The stochastic description therefore continues to capture non-trivial transitions between empirically distinguishable histories even as observational resolution is increased, despite the difference in how the dynamically discriminating information is represented. But the existence of deterministic and indeterministic alternatives does not depend on this feature. In dynamically stable regimes, fine-grained differences may remain small over relevant timescales, so that deterministic evolution provides the natural predictive description; yet stochastic alternatives that differ only within the unresolved fine structure can remain observationally equivalent at the given resolution, and empirical refinement may later discriminate between such alternatives.

If we grant that empirically established structure can be described on its own terms, then the usual view of the matter is something like this: there are generally various hypothetical completions available, positing rival ontologies; some such completion must ultimately be right; our inability to determine which is an epistemic matter; and while we press on with collecting further evidence, the best we can do is favour the alternative possessing the strongest scientific virtues---simplicity on Ockhamist grounds, explanatory economy, or the greatest promise of theoretical unification.

There is an implicit assumption here that our most fundamental theories resemble reality as it is in itself in a way that may not be justified. Take Newtonian gravitation. It does extremely well on simplicity and unification, but clearly breaks down in certain empirical domains. This is unproblematic because we know that Newtonian gravitation is not fundamental: reality is ``really'' quantum mechanical---or field-theoretical, or relativistic, or perhaps string-theoretical---and we have a reasonably good understanding of how Newtonian gravitation is recovered as an appropriate classical limit in at least some of these frameworks. If we are particularly open-minded, we may even admit that there could be something more fundamental that we have not yet thought of.

But when it comes to realist commitment, we nevertheless tend to identify reality with the entirety of some particular representational framework rather than with the structure that has actually been empirically established. Even when we speak of the ontology of theories such as Newtonian gravitation, which we already know to be idealised limits of more general descriptions, it is difficult to separate the full ontology posited by the theory---which, taken as universally valid, is clearly wrong---from the sense in which the theory nevertheless seems to give an exact description of certain relations obtaining in reality under the idealising conditions that define its regime of validity.

It seems clear, then, that there is a distinction to be drawn between structure that has been well established within a specified empirical domain and a space of alternative completions that may be well motivated scientifically, but which are far from unique in the additional structure they posit beyond the empirical horizon. Even where the present limits of empirical access are contingent rather than fundamental, it would seem reasonable to suspend realist commitment among competing completions---whether or not that space has been exhaustively characterised---and to accord primary ontological weight to the empirically corroborated structure that remains invariant across them.

Thus there \emph{may} be a fact of the matter as to whether reality is deterministic or indeterministic at the most fundamental level, and there is certainly a fact of the matter about whether a given deterministic or indeterministic model provides a good qualitative description of a given physical situation at a given finite level of resolution. But the redistribution of modal information involved in passing between deterministic and stochastic formulations beyond a fixed empirical horizon concerns precisely how the information selecting one of the empirically admissible histories is represented, rather than altering the empirically fixed space of those histories or their probability distribution: in one representation the selection is encoded in refined initial or boundary conditions and dynamics, while in the other it is distributed across stochastic outcomes. In neither case does the completion explain why this particular history rather than another member of the same weighted possibility space is realised; the selecting information is supplied to the representation rather than derived from the empirically corroborated structure. It is therefore possible to remain agnostic not only about which form of representation is more fundamental, but about whether the distinction between determinism and indeterminism per se corresponds to any fundamental physical fact at all.

 If attention is focused on the preparation--measurement correlation structure, it is possible to isolate structure present at a particular observational level, or range of such levels, while bracketing the possibility that the structure postulated by a given model breaks down at currently inaccessible resolutions. Classical Hamiltonian structure, for example, may be understood as a natural description of dynamical relations that are empirically corroborated across an accessible range of scales; taken to remain exact under arbitrary refinement, it extrapolates those relations to arbitrarily fine scales, but this continuation is an additional hypothesis, not required in order to grant the corroborated structure ontological weight within the observed regime. In this sense, a smooth Hamiltonian extension may constitute a particularly natural deterministic completion of empirically established structure, and may enjoy a privileged methodological status in directing further empirical inquiry. Conversely, where finite observational resolution yields stochastic transitions between empirically distinguishable states, the corresponding stochastic description may provide a more direct representation of the structure that is actually invariant at that level, without committing to the fine-grained continuation supplied by a Hamiltonian completion, or prejudging whether any deterministic completion whatsoever describes the fundamental ontology. The two representations therefore have different scientific virtues; neither virtue by itself settles the ontological question.

The argument of this paper is therefore that empirically corroborated invariant structure should be the primary locus of ontological commitment, while additional grounds are needed to privilege one representational formulation over another. Extending empirically corroborated structure to finer scales or different experimental setups may provide a useful scientific hypothesis, and considerations of simplicity, continuity or theoretical integration may give us reason to prefer one such extension over another. Where observations in an extended empirical regime corroborate additional structure, the equivalence class of admissible completions is correspondingly narrowed, and the ontological commitments licensed by the evidence may change. But until that extension is independently supported by the empirical record, it should be regarded as a hypothetical model of as-yet unprobed dynamics rather than as warranting ontological commitment at the refined or expanded level. Indeed, I will argue that, in the cases considered here, the freedom to choose among empirically admissible completions---including both deterministic and indeterministic formulations---constitutes a general gauge-like freedom\footnote{Here `gauge' denotes representational redundancy relative to a fixed empirical domain, rather than gauge symmetry in the usual field-theoretic sense. Below, there is a closer comparison with the broader literature on gauge freedom, invariance and representational surplus; see in particular Belot (2003), Møller-Nielsen (2017), and Dewar (2019).} in our representation of the underlying dynamics, one that can in many important cases be given a precise mathematical formulation. The analogy is with familiar gauge-theoretic cases in which distinct representatives preserve common invariant structure while differing in representational surplus; here, however, the equivalence is relative to a fixed empirical domain and may relate formally heterogeneous models as well as exact symmetry-related representatives. In the classical case, preservation of Liouville measure and associated finite-resolution dynamical structure provide important examples of the invariants that may survive across such alternative completions.

The proposal is not restricted to Werndl's cases in which deterministic and indeterministic formulations are comparably natural scientific models. Even where one representation is clearly more useful scientifically---whether because the observed regime is predictively stable or characteristically stochastic---alternative completions remain possible beyond the empirically established domain.

The model-invariant ontology advocated here locates ontological significance instead in the modal and relational structure encoded in preparation--measurement correlations: the weighted structure of empirically distinguishable possible histories and their relations, including transition probabilities, correlations, conservation constraints and symmetries. Particular deterministic or stochastic models may realise this structure through different microstates and different allocations of trajectory-selecting information; it is the structure common to the empirically admissible completions that receives primary ontological commitment.\footnote{Ontological commitment to this weighted structure does not by itself settle the interpretation of the weights as probabilities; whether they should additionally be understood as objective chances, propensities, epistemic quantities, or otherwise requires further argument.}

To make this proposal concrete, I use the Bernoulli and Baker maps as paradigm classical examples. The deterministic--stochastic equivalences involved are themselves familiar: any stochastic process can be represented deterministically on an appropriately defined path space\footnote{A Kolmogorov-style path-space formulation is a purely formal construction and need not be taken seriously as a physical alternative; Werndl describes it as a ``cheat,'' since its states are constructed to encode the entire history of the stochastic process. The significance of the construction here is just that stochasticity admits deterministic representation---less ``cheat-like'' alternatives may also be available. The stronger question is whether a deterministic formulation can independently earn realist consideration as a representation of physical structure---either because the relevant structure is empirically corroborated within a regime under consideration, or, more provisionally, because a refined empirical regime could in principle discriminate and support such a completion.}, with stochasticity carried by a probability measure over trajectories (Kolmogorov 1950; Doob 1953; Werndl 2009, 2011), while deterministic models with inaccessible fine-grained microstates routinely admit stochastic coarse-grained descriptions that suppress microstate detail while preserving observable behaviour (van Kampen 2007; Sklar 1993). The distinctive contribution of this paper is to use these representational dualities to formulate a model-invariance criterion for ontological commitment and to characterise the structure that remains invariant across compatible completions. A finite-resolution reconstruction of the Baker map illustrates how invariant probability-flow structure can be isolated from its deterministic or stochastic completion, while a two-level quantum system shows how an analogous distinction can be drawn between empirically fixed transition structure and the model-dependent representation of the dynamics that underlies it.

Whatever force one assigns to such constructions in the classical case, quantum mechanics provides an especially clear motivation for withholding ontological commitment from structure that is underdetermined by the empirical content of a theory. Here, formulations that agree on the empirically corroborated transition structure can differ radically in the additional ontology they posit and in whether the underlying dynamics is taken to be deterministic or indeterministic. The point is most relevant in quantum foundations, where otherwise sharply opposed realist programmes---for example, primitive-ontology approaches (e.g. Allori et al. 2008) and Everettian approaches (e.g. Wallace 2012)---can reproduce the same established quantum statistics while assigning radically different ontological significance to trajectories, branching, collapse, and the localisation of modal selection. Unlike in the classical examples, however, finer quantum empirical access generally requires a new measurement interaction and associated state transformation rather than merely a finer passive partition of a pre-existing state space. The empirically fixed quantum structure is therefore more naturally characterised as a relational structure connecting preparations, interventions, and outcomes, for which radically different ontological completions remain available.

My model-invariance criterion for ontological commitment proposes that within a specified empirical regime, ontological commitment should attach primarily to empirically corroborated features that remain invariant across the space of compatible, empirically equivalent formulations, or whose empirically accessible physical effects are preserved under such reformulations---in the examples considered in this paper, these features are principally modal and relational. This shifts attention away from the particular mathematical form of a model and towards the structure that persists through representational variation. The resulting position is a fallibilist and regime-relative form of structural realism: it permits realist commitment to invariant structure without treating any presently successful model as an exhaustive description of the underlying ontology, and leaves open the possibility that finer empirical access may disclose additional structure or revise existing commitments. Determinism and indeterminism, insofar as they vary across compatible completions while leaving the empirically corroborated modal structure unchanged, accordingly lack independent ontological significance within the corresponding domain. This does not imply that relations exhaust ontology: objects and properties may equally warrant commitment where their identity and relevant properties are themselves invariant across admissible representations.

Thus, in contrast with both ontic structural realism, which treats relational structure as exhaustive of ontology (e.g. Ladyman \& Ross 2007), and epistemic structural realism, which restricts realist commitment to the structural content that survives theory change while remaining agnostic about the nature of the underlying entities (e.g. Worrall 1989), the present view takes satisfaction of the model-invariance criterion to provide positive, though defeasible, warrant for ontological commitment.

\section{Model-equivalence: illustrative examples}

\subsection{The Bernoulli map}

To illustrate deterministic--stochastic model-equivalence in a maximally transparent way, consider the Bernoulli map,
\[
T(x)=2x\bmod 1,\qquad x\in[0,1).
\]
This map doubles the value of \(x\) and shifts it back into the unit interval; it is a standard example of a chaotic dynamical system. Iterating the transformation produces a sequence of values determined completely by the initial point. If the interval is partitioned into the two halves \([0,1/2)\) and \([1/2,1)\), and we record at each iteration which half contains the point, we obtain a binary sequence. For an initial point distributed according to Lebesgue measure, this sequence has exactly the probability distribution of an independent sequence of fair coin tosses.

Thus the Bernoulli map, although strictly deterministic, generates the same observable symbolic dynamics as a Bernoulli stochastic process when examined through this partition. At this observational level the two descriptions have the same empirical content: the deterministic formulation represents the entire symbolic history as fixed by the fine-grained initial state, while the stochastic formulation represents successive symbols probabilistically. Conversely, the stochastic process itself admits a deterministic path-space representation in which the realised sequence is encoded in the initial point. What is common to these representations is the weighted space of empirically distinguishable histories---the possible binary sequences together with their probability distribution---rather than the particular way in which selection of one history is represented.

The coarse partition should be understood as specifying an empirical regime, not as asserting a fundamental limit on preparation or measurement. At that level, underlying points that generate the same accessible symbolic information are empirically indistinguishable, and nothing in the observed dynamics determines whether the unresolved structure should be interpreted literally as fine-grained spatial position or in some other way. Finer empirical access may subsequently distinguish between completions that were equivalent under the original partition; if so, the empirical regime is refined and the model-invariance class correspondingly narrowed.

To understand this distinction, consider the deterministic representation. Taken as a literal representation of fine-grained physical state, the Bernoulli map posits a state specification that, together with the dynamics, fixes the complete symbolic history. The model may indeed remain valid under arbitrary refinement, but its empirical success at the coarse level does not by itself warrant that extrapolation. To the extent that the fine-grained structure remains unresolved or, depending on the physical context, does not correspond to a physical quantity obviously susceptible to physical probing at that level of refinement, the representation may instead be regarded as bookkeeping for the space of possible histories. Viewed measure-theoretically, this change of interpretation need make no difference to the weighting of those histories, provided that the measure over history space is preserved.

Calling the unresolved structure ``bookkeeping'' is not intended dismissively. A completion must represent somehow the information selecting which admissible history is realised. In the deterministic formulation this information is encoded in the fine-grained initial condition; in the stochastic formulation it is distributed across successive random outcomes. Neither representation thereby explains why this particular history rather than another member of the same weighted possibility space obtains. What differs is the localisation and modal character of the selecting information, not the empirically fixed space of coarse-grained histories or its probability distribution.

The stochastic representation therefore displays the model-invariant structure particularly directly, without requiring that its stochasticity be interpreted as fundamental. The deterministic completion, by contrast, supplies additional fine-grained structure and \emph{may} thereby suggest empirical tests capable of breaking the present equivalence. In the present toy model, such refinement is naturally represented as increased spatial resolution, but this depends on the physical interpretation of the model: the additional structure may correspond to quite different degrees of freedom, or may not be directly accessible to physical probes at all. The distinction is not between scientifically legitimate and illegitimate representations, but between structure already corroborated within the specified regime and completion-specific structure whose further ontological warrant remains open.

The Bernoulli map gives the simplest possible example of this distinction. In the more general classical setting developed below, the idea can be developed by asking which finite-resolution dynamical structures survive across alternative completions. For Hamiltonian systems, preservation of Liouville measure provides one such constraint: distinct fine-grained realisations may preserve the same empirically accessible probability flow while differing in their underlying microdynamics.

\subsection{Effective Dynamics and Gauge Invariance}

To begin to generalise the above example, consider the case where we can improve our observational resolution of the Bernoulli map (imagining that it is applied as a model to some real-world situation) to an arbitrary finite extent. Suppose we can increase the resolution of our apparatus to gain empirical access to a more precise, but not exact, value of the initial condition and/or the subsequent values instantiated by the system. At any such finite resolution, the deterministic system, together with the observation function defined by the chosen partition and the invariant Lebesgue measure on \([0,1)\), induces a corresponding stochastic process on the empirically distinguishable states. Our epistemic situation is therefore not qualitatively different: the deterministic model remains empirically indistinguishable, relative to that observational scheme, from the stochastic process so induced. Our improved knowledge may allow us to predict certain aspects of the coarse-grained behaviour for a finite number of forward steps, but the transitions between kinematically distinguishable positions again follow a well-defined and non-trivial stochastic process. For example, if we divide the unit interval into cells of length \(1/2^n\) and possess an apparatus that allows us to distinguish which of these cells contains the value of \(x\) at any given step, then we will now be able to predict the left/right position for \(n-1\) steps of the process. Alternatively, for these \(n-1\) steps, we retain some information about which of the \(2^n\) refined cells the system can occupy. Even in the stochastic formulation, the dyadic dynamics plays a mathematically specifiable geometric role encoded exactly in the transition probabilities. The kinematically distinguishable states---namely, the distinguishable positions defined by the refined partition---correspond to initial sequences of length \(n\) in the binary expansion of \(x\), and the induced process is Markovian on these configurational states. Once the current refined state is specified, the past is completely screened off: the transition probabilities depend only on that state.

If, perhaps more realistically, our partition does not fit neatly into the structure governed by the underlying dynamics---for example, if we divide the interval into the two cells \([0,1/3)\) and \([1/3,1)\)---the symbolic process induced directly by that partition will not, in general, be Markovian. Nevertheless, the process can be approximated to arbitrary finite observational precision by taking a sufficiently fine dyadic partition as the underlying state space: unions of dyadic cells can approximate the non-dyadic observational cells arbitrarily closely in measure, while the dynamics on the refined dyadic states remains Markovian. The non-Markovian process associated with the coarser observation can therefore be represented, to any required tolerance, as the coarse observation of a refined Markov process. This provides a concrete instance of the general \(\varepsilon\)-congruence results discussed by Werndl (2009) and Ornstein and Weiss (1991). More
generally, whenever preparational states do not encode sufficient information
to fully characterise future evolution---either because the partition is not
dynamically aligned or because states are represented by distributions over
cells---the induced one-step dynamics will typically be non-Markovian. 

In what follows, the discussion will mostly be confined to dynamically aligned cells that expose a particularly simple transition structure. Such alignment need not be regarded as entirely artificial: where it corresponds to empirically observable simplifications in the transition structure, it may well reflect real dynamical structure. However, it should be borne in mind that a general representational framework cannot assume that every empirically natural partition will have this property.

The Bernoulli example illustrates a distinction that will be important throughout what follows. The fine-grained deterministic dynamics may be retained as a substantive hypothesis about presently unresolved structure, but the ontological warrant supplied by the empirical evidence attaches more securely to the modal and dynamical structure that is preserved at the observational level. Representations that agree on this corroborated structure may nevertheless differ substantially in how they complete it beyond any empirical resolution currently available.

At a given finite resolution, the coarse-grained symbolic dynamics reproduces the empirically accessible evolution while remaining neutral with respect to distinctions among fine-grained states that make no difference at that level. Quotienting the fine-grained state space by this observational equivalence therefore introduces a representational freedom: different organisations of the unresolved microstructure, and in some cases different underlying dynamics, may induce the same transition structure on the empirically distinguishable states. It is this freedom that I will treat as gauge-like. The next question is which features of the underlying dynamical structure are preserved across such alternative completions, and whether those invariants can be supplied with a convenient mathematical description directly at the coarse-grained level.

We should note that the relevant equivalence classes need not be defined solely by instantaneous kinematic indistinguishability. Where the coarse-grained dynamics retains history dependence, empirical states are more naturally individuated by the observable probability distributions they induce over possible future histories: underlying states that generate the same accessible predictive structure are empirically equivalent even if their fine-grained representations differ. Dynamically aligned partitions are a special case in which this predictive structure closes particularly simply on the instantaneous coarse-grained states themselves.

To make the relation between unresolved sub-cell structure and observable history dependence a bit more explicit, consider the two-cell partition mentioned above
\[
A=[0,1/3), \qquad B=[1/3,1).
\]
With respect to the invariant Lebesgue measure, the symbolic process on \(A\) and \(B\) is not Markovian. If the present state \(B\) was reached from \(A\), the induced distribution over fine-grained positions within \(B\) differs from that obtained when \(B\) was reached from \(B\). Consequently,
\[
P(A_{t+1}\mid B_t,A_{t-1})=\frac12,
\qquad
P(A_{t+1}\mid B_t,B_{t-1})=\frac16.
\]
The deterministic representation retains the sub-cell positional information that distinguishes these cases; once this information is suppressed by the coarse-graining, it reappears as dependence on the observed history. What is represented as hidden fibre information in the fine-grained description can therefore be represented as non-Markovian correlation in the coarse-grained stochastic description.

This also suggests distinct empirical strategies for probing the underlying invariant structure. One may increase positional resolution and attempt to resolve the sub-cell distinctions directly, or one may remain at the same kinematic resolution while measuring correlations across progressively longer histories. Whether a refined or enlarged state description yields a simpler, approximately Markovian transition structure is itself an empirically testable feature of the dynamics. These strategies therefore provide complementary ways of investigating how far the established dynamical regularities persist as the empirical domain is extended.

The fine-grained map thus retains a legitimate constructive role as a possible extension of the empirically corroborated dynamics. By contrast, the coarse-grained and stochastic representations help to isolate the modal and dynamical structure that survives the redistribution of information between fine-grained states and observable histories. They do not establish that stochasticity is more fundamental, but they make the model-invariant structure particularly explicit.

\subsection{Reversibility}

Lest one suspect that the preceding discussion depends upon the inherently
irreversible character of the Bernoulli map, it is natural to turn next to its
simplest invertible generalisation, the Baker’s map. This reversible system also
provides a first indication of how the gauge-like representational freedom alluded to in the previous
section manifests in the more general setting of symplectic invariance and
Hamiltonian flows.

The Baker’s map $B\colon[0,1)\times[0,1)\to[0,1)\times[0,1)$ is defined by\footnote{The dyadic rationals $x=k/2^n$ have non-unique binary expansions, and
are sometimes excluded from the domain. As they form a zero-measure set, this
does not affect anything either way at the level of the coarse-grained theory, whose partitions are of finite measure. The map as defined
here is bijective.}
\[
B(x,y)=
\begin{cases}
(2x,\, y/2), & 0 \le x < 1/2, \\
(2x-1,\,(y+1)/2), & 1/2 \le x < 1.
\end{cases}
\]
This map is area-preserving and, because it is invertible, every point on a
trajectory encodes the full past and future behaviour. The projection onto the
$x$-coordinate reproduces the Bernoulli map, but the encoded trajectory
information that previously appeared to be carried solely by the initial
condition can now be localised at any point, or even distributed across the
entire trajectory. This is, of course, a general feature of reversible dynamical
systems, yet it underlines the fact that trajectory information underdetermined
by the dynamics cannot be given a privileged temporal location even in the
deterministic case. As before, coarse-graining the system by dividing the
kinematic state space into finite cells introduces a gauge-like freedom: microstates
(and, if we wish, microdynamics) can be reorganised in any way that preserves the
modalities of the coarse-grained theory. The resulting coarse-grained dynamics can accordingly be represented by a suitable stochastic process---a two-sided Markov shift for dynamically aligned partitions, or more generally by a process admitting arbitrarily close Markov approximation in the sense of \(\varepsilon\)-congruence. The deterministic--stochastic distinction therefore remains representational at the coarse-grained level for the same reason as in the Bernoulli case: the alternative formulations preserve the same empirically accessible modal structure. Reversibility just adds a slight twist about localisation, since no point on the trajectory is temporally privileged as the unique bearer of trajectory-selecting information.

It is worth emphasising that the stretching--and--stacking mechanism embodied in the Baker’s map is not as ``artificial'' as Werndl’s description of it might suggest. The map is idealised, but it provides a particularly simple representation of a dynamical pattern that occurs widely in physical models: local expansion in one direction and contraction in another, together with folding or reinjection that keeps the evolving structure within a bounded region. This occurs, for example, in return maps for chaotic Hamiltonian systems and in various models of fluid mixing. The Baker’s map thus isolates in analytically transparent form a mechanism that more complicated physical dynamics often instantiate.

The connection to Hamiltonian dynamics can be made particularly transparent in this model. The area-preserving property of the Baker’s map is a two-dimensional instance of symplectic invariance: away from the discontinuity, each branch preserves the non-degenerate closed 2-form
\[
\omega=dx\wedge dy.
\]
A convenient Hamiltonian for the stretch--contract mechanism is
\[
H(x,y)=xy.
\]
Hamilton’s equations are
\[
\dot{x}=x,\qquad \dot{y}=-y,
\]
whose exact flow over a time \(\ln2\) is
\[
(x,y)\longmapsto(2x,y/2).
\]
This is the first branch of the Baker’s map. The second branch differs from it only by a translation,
\[
(2x,y/2)\longmapsto(2x-1,(y+1)/2),
\]
which also preserves the symplectic form. The Baker’s map may therefore be regarded as a piecewise-affine symplectic transformation: each branch preserves \(\omega\), and hence the associated Liouville measure, while the cutting and stacking operation makes the global map discontinuous. There is a slight change of perspective here, in that \(x\) and \(y\) are now interpreted as canonical phase-space coordinates rather than simply as configuration-space coordinates, but this makes sense formally.

Coarse-graining destroys the one-to-one correspondence between empirically distinguishable states even though the underlying map remains bijective. Let \(\{C_i\}\) be a finite partition of the phase space, and let \(\mu\) denote the invariant Liouville measure. The induced one-step transition probabilities are
\[
P_{ij}
=
\frac{\mu(C_i\cap B^{-1}(C_j))}{\mu(C_i)}.
\]
Because \(B\) preserves \(\mu\),
\[
\sum_i\mu(C_i)P_{ij}=\mu(C_j),
\]
so the measure of the cells defines a stationary distribution for the coarse-grained dynamics. For an equal-area partition, such as a partition into dyadic squares, this distribution is uniform and the transition matrix is bistochastic:
\[
\sum_jP_{ij}=\sum_iP_{ij}=1.
\]
For dynamically aligned partitions this transition matrix may determine a Markov process directly; for more general partitions it describes only the one-step transition structure, with additional history dependence retained in higher-order conditional probabilities.

This gives a precise sense in which a consequence of symplectic invariance survives coarse-graining. What is preserved is not the full fine-grained symplectic geometry or microscopic reversibility, but the Liouville measure and the corresponding probability-flow constraints on the empirically distinguishable states. At the level of individual microstates the Baker’s map remains fully reversible; what is lost is the ability of the coarse-grained state to specify a unique microscopic successor or predecessor. Distinct microhistories consequently become observationally indistinguishable, and the effective dynamics may be represented stochastically.

The stochastic description is therefore not an ad hoc replacement for the Hamiltonian dynamics. It expresses at finite resolution structure inherited from the measure-preserving fine-grained model while remaining neutral about unresolved microstructure. In particular, the loss of effective reversibility under coarse-graining does not establish an underlying irreversible ontology. What survives across the deterministic and stochastic representations is the invariant measure and associated transition structure; the particular localisation of the information required to realise individual trajectories does not.

One might object that where the stochastic transition structure is derived from a deterministic completion, the latter should thereby receive ontological priority. But derivational priority within a model does not by itself establish ontological priority. Scientific models are idealising constructions: in isolating particular structural relations from background effects and unresolved degrees of freedom, they are warranted only relative to the regimes in which those idealisations are empirically successful. The fact that a stochastic structure can be obtained by coarse-graining a deterministic model therefore does not by itself establish that the fine-grained deterministic structure remains ontically valid beyond the regime in which it is corroborated.

The general lesson here is that coarse-graining a Hamiltonian flow by a finite partition induces a stochastic process on empirically distinguishable states, which need not in general be Markovian. Although full symplectic geometry is not preserved at the coarse-grained level, its discrete signature, Liouville-measure preservation, constrains the induced probability flow, and for equal-measure cells appears directly as bistochasticity of the one-step transition matrix. What survives is not a particular fine-grained dynamical law, but a finite-resolution measure and transition structure inherited from it.

Distinct deterministic, stochastic or hybrid completions may realise this same empirically corroborated structure while differing in how they represent unresolved dynamics. They are gauge-like alternatives relative to the fixed empirical domain: at the coarse-grained level they agree on the ontological structure warranted by the evidence, while at finer levels they may imply rival ontologies between which the available evidence does not yet justify a choice. The stochastic representation, however, is particularly useful in the sense that it displays the invariant probability structure directly, without needing to imply either fundamental stochasticity or a commitment to any particular fine-grained completion.

\subsection{Quantum two-level system}

Model invariance in quantum mechanics may be illustrated by considering a
two-level system (qubit) evolving under a fixed Hamiltonian (e.g.\ a uniform
magnetic field) and subjected to a coarse-grained measurement described by a
positive operator-valued measure (POVM).
The aim at this stage is not to revisit foundational questions about quantum measurement, but to show explicitly how the choice of a POVM defines an observational coarse-graining of quantum state space, identifying states that give the same accessible outcome statistics. When the underlying unitary dynamics is projected through this many-to-one observational map---with state update specified by an appropriate instrument---the resulting empirical dynamics is stochastic.

The state of a two-level quantum system is represented by a density operator
\[
\rho \in \mathcal{D}(\mathbb{C}^2), \qquad \rho \ge 0, \qquad \mathrm{Tr}\,\rho = 1.
\]
Any such state admits a Bloch representation
\[
\rho = \tfrac{1}{2}(I + \mathbf{r}\cdot\boldsymbol{\sigma}), \qquad
\mathbf{r} \in \mathbb{R}^3, \qquad \|\mathbf{r}\| \le 1,
\]
with pure states corresponding to points on the unit sphere
$\|\mathbf{r}\|=1$ and mixed states to interior points. The Bloch sphere thus
provides a convenient geometric representation of the quantum state space,
though it should be emphasised that it is a space of equivalence classes of
preparation procedures, not a classical phase space of simultaneously measurable
quantities.

Let the system evolve under a time-independent Hamiltonian
\[
H = \tfrac{\hbar\omega}{2}\,\mathbf{n}\cdot\boldsymbol{\sigma},
\]
where $\mathbf{n}$ is a fixed unit vector. The unitary evolution
\[
\rho(t) = U(t)\rho(0)U^{\dagger}(t), \qquad
U(t)=e^{-\,iHt/\hbar},
\]
induces a rigid rotation or precession of the Bloch vector $\mathbf{r}$ about
the axis $\mathbf{n}$ with angular velocity $\omega$. The dynamics of the model is therefore deterministic and invertible; on each fixed-spectrum orbit it admits a Hamiltonian formulation.

In this model, the density matrices evolve in closed orbits under the unitary
dynamics, and it is tempting to identify these orbits with ``trajectories'',
especially since density matrices provide a natural representation of the
correlation structure between preparations and observables that I have argued
constitutes the appropriate content of ontic commitment. It is important,
however, to remember that quantum theory does not provide empirical access to
anything corresponding to the classical notion of a trajectory. There are no
well-defined simultaneous values of non-commuting observables, and no
operational access to a continuous history of intermediate states. Any attempt
to probe the state at intermediate times introduces a measurement interaction and an associated state update into the description of the system and thereby defines a new dynamical situation. This is not
a limitation arising from experimental imprecision, but a structural feature of
the quantum description itself.

To make this explicit, consider a two-outcome POVM defined by the effects
\[
E_{\pm} = \tfrac{1}{2}(I \pm \eta\,\sigma_z), \qquad 0 < \eta \le 1.
\]
For any state $\rho$, the probabilities of the two outcomes are
\[
p(\pm\mid\rho) = \mathrm{Tr}(E_{\pm}\rho)
               = \tfrac{1}{2}(1 \pm \eta\, r_z),
\]
where $r_z$ is the $z$-component of the Bloch vector. The parameter \(\eta\) controls the sharpness of the measurement: \(\eta=1\) corresponds to a projective measurement of \(\sigma_z\), while smaller values of \(\eta\) define increasingly unsharp observational probes, analogous to progressively coarser partitions in the classical phase-space examples.

This POVM defines a many-to-one observational map from the space of density
operators to two outcome probabilities. States that differ in their transverse
components $(r_x,r_y)$ but agree in $r_z$ are empirically indistinguishable at
this level of description.

If one now seeks a description formulated entirely at the level of these coarse-grained outcomes, an additional ingredient is required. A POVM fixes the probabilities of the possible outcomes but does not uniquely determine the state update associated with them. Let \(\{\mathcal I_i\}\) be an instrument compatible with the effects \(E_i\), so that
\[
p_i=\operatorname{Tr}\mathcal I_i(\rho)=\operatorname{Tr}(E_i\rho),
\]
and, conditional on outcome \(i\),
\[
\rho_i'=\frac{\mathcal I_i(\rho)}{p_i}.
\]
After a further interval of unitary evolution, the conditional probability of outcome \(j\) is therefore
\[
p(j\mid i,\rho)
=
\operatorname{Tr}\!\left(
E_j U(\Delta t)\rho_i'U^\dagger(\Delta t)
\right).
\]
Whether this closes to a transition matrix defined purely on the outcome labels depends on how much information about the incoming state is retained by the instrument.

For an ``outcome-resetting'' instrument---that is, one in which the post-measurement state is fixed entirely by the observed outcome---we may write,
\[
\mathcal I_i(\rho)=\operatorname{Tr}(E_i\rho)\tau_i,
\]
where \(\tau_i\) is a fixed density operator associated with outcome \(i\). Since \(\tau_i\) is fixed by the outcome, the post-measurement state carries no information about the incoming state beyond \(i\). The outcome dynamics then closes as a Markov process with
\[
T_{ji}
=
\operatorname{Tr}\!\left(
E_jU(\Delta t)\tau_iU^\dagger(\Delta t)
\right).
\]
At the other extreme, for a minimally disturbing Lüders instrument, taking
\[
\mathcal I_i^{L}(\rho)=\sqrt{E_i}\rho\sqrt{E_i},
\]
an unsharp measurement with \(0<\eta<1\) preserves part of the state information not encoded in the observed outcome. In Bloch form, for outcome \(s=\pm1\),
\[
r_x'=\frac{\sqrt{1-\eta^2}\,r_x}{1+s\eta r_z},
\qquad
r_y'=\frac{\sqrt{1-\eta^2}\,r_y}{1+s\eta r_z},
\qquad
r_z'=\frac{r_z+s\eta}{1+s\eta r_z}.
\]
Thus the transverse components are attenuated but not in general erased. If the subsequent Hamiltonian evolution mixes these components with the \(z\)-direction, they can affect later outcome probabilities even though they are not themselves resolved by the POVM. The present outcome then need not screen off the previous history, and the induced process on the outcome labels may be non-Markovian. Closure can be restored by enlarging the effective state description to retain the relevant information.

This reproduces closely the distinction encountered in the classical examples. There, unresolved position within a coarse cell may either be incorporated into a refined state description or reappear as correlations across the observed history. Here, information retained in the post-measurement quantum state but omitted from the POVM outcome plays the corresponding role. The choice of instrument therefore determines how much of the unresolved structure is carried forward in the state and can subsequently reappear as temporal dependence in the observable process.

The stochasticity is nevertheless not added by hand. The POVM defines a many-to-one observational coarse-graining of the quantum state space, while the instrument specifies how much structure outside that observational content survives the intervention. Different instruments implementing the same POVM agree on the immediate outcome statistics but can differ in their predictions for subsequent or finer-grained measurements. They are therefore gauge-like alternatives only relative to the restricted empirical domain fixed by the POVM; once sequential measurements probe their differing state updates, the empirical equivalence class is correspondingly refined. This freedom should also be distinguished from the genuine representational redundancy among different Kraus decompositions of one and the same instrument or channel.

The structural parallel with the classical examples is now clear: in both cases a many-to-one observational coarse-graining fails to resolve all dynamically relevant information, and whether the effective dynamics closes on the instantaneous observational states depends on how much of that information is retained in the underlying state description.

The quantum case reformulates the general lesson by making the role of observation explicit and dynamically relevant. The adoption of a POVM does not merely specify which aspects of the state are empirically resolved; it also constrains the possible state updates, and hence the character of the effective dynamics at that observational level. The resulting operational description consists of unitary evolution---understood as part of a fine-grained representation---interspersed with probabilistic transitions associated with observation. At the chosen observational resolution, what is empirically accessible is therefore a stochastic process over coarse-grained outcomes, irrespective of whether some finer-grained representation of the underlying dynamics is deterministic.

The important difference from the classical examples is thus that the observational intervention is itself generally dynamical: a measurement is represented not only by probabilities for possible outcomes, but by an instrument specifying how the measurement interaction transforms the state. This transformation cannot in general be reduced to an epistemic update concerning an independently evolving underlying state. The relation between empirical access and effective dynamics is therefore more complicated than in the classical coarse-graining examples, though this by no means settles whether the underlying ontology is deterministic or indeterministic.

Note that an even closer parallel with the classical case could be preserved by coarse-graining a suitable quantum phase-space representation and postponing an explicit consideration of measurement. In such a formulation one could again identify a gauge-like representational freedom among descriptions that agree on the empirically accessible coarse-grained dynamics. The present POVM construction has the advantage of making explicit the dynamically invasive role that measurement can play in the quantum case.

Though coarse-graining induced by POVM measurement is only one of many possible coarse-graining types, the present example illustrates that the probabilistic outcome structure associated with invasive measurement does not, by itself, determine the ontological character of the underlying dynamics. It constrains the class of admissible descriptions, but leaves open how the dynamics is to be represented at levels not fixed by the observational structure.

Once a particular observational coarse-graining is fixed, the quantum formalism fixes a probabilistic preparation--outcome structure at the chosen observational level, while leaving open how information outside that observational content is represented and propagated. Different compatible instruments may retain different amounts of such information and thereby generate different sequential dynamics, just as alternative fine-grained completions may differ in their account of the underlying evolution. What is empirically fixed is therefore not a unique deterministic or stochastic ontology, but the modal structure common to the admissible representations at the specified observational level. A modest stance that remains agnostic about structure beyond the empirical horizon therefore retains the possibility of a mathematically well-defined structural description of the invariant empirical facts.

\section{General applications}

\subsection{Continuous time classical Hamiltonian dynamics}

Having considered simple examples of deterministic--stochastic model-equival-ence, we can now ask how generally the idea applies. In the classical case, the most important existing results are those of Ornstein and Weiss (1991) and Werndl (2009, 2011). The discussion below will focus on systems whose empirically accessible dynamics is compatible with a measure-preserving phase-space description; this concerns the structure exhibited at the empirical level and does not privilege any particular deterministic or stochastic completion of it.

Ornstein and Weiss establish at the level of rigorous ergodic theory that the distinction between deterministic and stochastic descriptions need not be invariant under observationally relevant reformulation. Their results concern in particular Bernoulli systems, whose measure-theoretic structure is classified up to isomorphism by Kolmogorov--Sinai entropy. In both discrete- and continuous-time settings, such deterministic systems can be matched arbitrarily closely by stochastic processes: every continuous Bernoulli system, for example, is \(\varepsilon\)-congruent, for arbitrary \(\varepsilon>0\), to a semi-Markov process and can be simulated to arbitrary observational accuracy by suitable \(n\)-step semi-Markov processes. Deterministic and stochastic formulations may therefore differ radically in mathematical form while agreeing to arbitrary finite empirical resolution.

Werndl sharpens and extends this result by bringing it explicitly into contact with scientific modelling practice. She distinguishes several notions of observational equivalence at every observation level and shows, for important classes of systems, that deterministic models found in science can be matched by stochastic models found in science to arbitrary observational accuracy, including under arbitrary finite-valued observation functions (Werndl 2009, 2011). At the same time, she suggests that observational equivalence alone does not establish that determinism and indeterminism are always mere modelling artefacts; wider theoretical and indirect evidential considerations may favour one model over another (Werndl 2013).

The present question is somewhat different. What matters is not whether competing formulations are equally ``science-like'', but whether they preserve the same empirically corroborated dynamical and probabilistic structure within the specified domain. Werndl's cases are especially symmetric, but model invariance does not require this symmetry. A stochastic representation may be scientifically preferable where it gives the most economical description of accessible history probabilities, while a deterministic representation may be preferable in a predictively stable regime. In either case, scientific utility within the specified empirical regime does not by itself settle the ontological status of completion-specific structure. The relevant distinction is between the empirically corroborated probability structure and the additional means by which a particular representation encodes selection of an actual history.

At the level of the idealised, dynamically aligned dyadic partitions of the Baker's map, the invariant structure is particularly simple: Liouville-measure preservation induces a well-defined probability flow between coarse-grained states, and for equal-measure cells this appears as a bistochastic transition matrix. More general observational coarse-grainings need not exhibit this simple Markovian structure. Dynamically relevant information omitted from the instantaneous coarse state may instead reappear as correlations across observable histories. The model-invariant structure is therefore not any particular transition matrix, but the measure and probability structure over empirically distinguishable histories preserved across admissible representations.

At this point, if we are going to pursue the idea of observational equivalence as ``gauge-like,’’ it is useful to make some comparisons between what we are doing and the extensive literature on gauge freedom and representational surplus. A recurrent methodological idea across much of the literature on gauge symmetry is that invariant structure and features that vary between symmetry-related representatives should be accorded a different ontological status. The strongest versions of this principle identify physical reality exclusively with gauge-invariant structure, but this inference has been contested. M{\o}ller-Nielsen (2017), for example, argues that symmetry-variance should motivate suspension of ontological commitment rather than being taken as sufficient evidence of unreality. This position provides a precedent for the fallibilist distinction proposed here: structure that varies across empirically equivalent representations need not be dismissed as unreal, but its ontological status remains less securely warranted than that of empirically corroborated invariant structure and may be revised when empirical refinement breaks the current equivalence class.

The wider literature on theoretical equivalence distinguishes empirical equivalence from stronger formal and interpretational notions of equivalence (Weatherall 2019a,b). Within the literature on gauge freedom, at least two further questions are particularly relevant here. The first is whether symmetry-related models should themselves be regarded as representing the same physical state of affairs. Interpretationalist approaches take the symmetry to license such an identification, whereas motivationalist approaches treat it only as a reason to seek a more perspicuous account of the common ontology (M{\o}ller-Nielsen 2017; Martens and Read 2021). The second question is how that common content should be represented. Reductive approaches seek a formalism containing only gauge-invariant structure (Belot 2003), while ``sophisticated’’ approaches retain the richer representatives but reinterpret the symmetry transformations as expressing physical equivalence (Dewar 2019). Related work has also emphasised that representational surplus need not be scientifically superfluous: redundant structure may contribute essentially to locality, explanatory practice, or the representation of a rich space of possibilities (Nguyen, Teh, and Wells 2020; Bradley and Weatherall 2020).

Of course, the claim here is not that model invariance is identical with gauge invariance, nor that empirical equivalence entails theoretical or interpretational equivalence in the stronger senses distinguished by Weatherall. Empirical equivalence need not amount to theoretical equivalence, and empirically equivalent formulations may differ in the amount or character of structure they posit. Clearly, empirically equivalent deterministic and indeterministic formulations, as commonly construed, postulate rival ontologies in a way that different gauge representatives are often considered not to---though interpretations on which gauge-related models represent distinct physical possibilities remain available. The important difference for present purposes, however, is that model-invariant structure, while warranting ontological commitment, is explicitly not intended to provide an exhaustive ontological account. The ontological commitment to such structure remains fallibilist in at least two distinct senses: it is open to the possibility of misidentification within a specified empirical regime, and it may be differently contextualised when further empirical evidence from previously untested regimes becomes available. But the idea is that there is a strong case that the empirically corroborated structure isolated by this criterion is genuinely ontological and may, if necessary, be furnished with a self-standing mathematical description. My argument is that this places it in a different epistemic category from structure whose warrant remains specific to one or more variant representatives of the model-invariance class, however compelling the indirect evidence for favouring such structure as a possible extension of the invariant ontology.

Framed in this way, the present analysis naturally aligns with the ``reverse physics'' programme developed by Carcassi and Aidala (2022). Rather than taking a familiar mathematical formalism as fundamental, this programme asks which physically meaningful assumptions are sufficient to reconstruct it. In their treatment of classical mechanics, experimentally distinguishable states are represented by a topological state space, deterministic and reversible evolution by structure-preserving transformations of that space, and the assumption of infinitesimal reducibility supplies the further structure from which classical Hamiltonian mechanics can be recovered. Hamiltonian structure is therefore obtained from a relatively abstract package of assumptions concerning states, their empirical distinguishability, composition and evolution, rather than from commitment to any particular microscopic ontology.

This provides a useful perspective on the role of Hamiltonian mechanics in the present argument. Where the relevant dynamical relations remain stable across an empirically accessible range of scales, their smooth continuation under arbitrary refinement provides a particularly natural deterministic completion of that structure. Nothing in the success of this completion within the observed regime, however, requires that the same structure remain exact at indefinitely finer scales. The Hamiltonian representation may therefore capture genuine model-invariant structure within its domain of empirical support even if further refinement ultimately reveals dynamics that require a different ontology or formalism. This possibility is illustrated within Carcassi et al.'s reconstruction of Hamiltonian dynamics (2018): replacing the assumption of infinitesimal reducibility with irreducibility leads not to the abandonment of Hamiltonian structure, but to a quantum Hamiltonian system.

Coarse-graining is the complementary side of this picture. A many-to-one identification of empirically indistinguishable states will in general induce a stochastic process over the resulting observational states, even where the fine-grained evolution is deterministic. In the Baker’s-map example, Liouville-measure preservation descends to an invariant probability measure on the coarse states and, for suitable equal-measure partitions, to a bistochastic one-step transition structure; more generally, it induces a measure over observable histories, including higher-order conditional probabilities and possibly non-Markovian temporal correlations.
This stochastic structure does not by itself determine a unique Hamiltonian realisation. Where its dynamical features remain stable across an empirically accessible range of scales, however, a smooth Hamiltonian completion may provide a particularly natural extrapolation of them to finer resolutions, without implying that this extrapolation must remain exact under every further empirical refinement.

The reverse-physics question can therefore be posed at the level of model-invariant structure itself. Given an empirically established state, measure and transition structure, what additional physically motivated conditions are sufficient to warrant a stronger dynamical representation---for example, a deterministic and reversible Hamiltonian realisation? As previously noted, this may provide a useful formulation of model-invariant structure even when it remains indexed to a specified empirical regime. Where those stronger conditions are not warranted even in a regime-relative sense, the stochastic transition or history structure can instead be treated on its own terms without prejudging which deterministic, stochastic or hybrid completion realises it in fact. In this sense, reverse physics provides a general methodology for asking how much theoretical structure is warranted by the physical relations established at a given empirical level.

\subsection{A model-invariance criterion for ontological commitment}

The analyses of the preceding sections motivate a general criterion for distinguishing between structural features of a theory that warrant ontological commitment and features that can vary under representational choice. The need for such a criterion arises whenever the empirically corroborated structure within a specified domain admits multiple formally or ontologically distinct formulations---deterministic and stochastic, Markovian and non-Markovian, Hamiltonian and non-Hamiltonian---that nevertheless agree on everything empirically fixed within that domain.

The guiding idea of the present framework is that ontological commitment should attach neither indiscriminately to individual models nor to their mathematical apparatus as a whole, but to the empirically corroborated structure that remains invariant across the admissible representations. Empirical accessibility is therefore always indexed to a specified regime \(R\), including the relevant resolution, timescale, preparations, interventions and observational scheme. Structure that is invariant relative to one such regime may cease to be invariant when that regime is extended, but this does not undermine its ontological warrant as a description of structure within the original regime.

More formally, let \(\mathcal{C}_{R}\) denote the equivalence class of admissible completions that reproduce the same empirically accessible preparation--measurement and history structure within a specified empirical regime \(R\), where \(R\) includes the relevant observational resolution and tolerance. A structural feature \(F\) warrants model-invariant ontological commitment within \(R\) only insofar as:
\begin{enumerate}
\item the structure associated with \(F\) is empirically corroborated within \(R\);
\item that structure is preserved across \(\mathcal{C}_R\), either directly or through invariant empirically accessible physical effects; and
\item the empirically corroborated invariant structure can be identified either as a standalone mathematical structure or as embedded in particular representations indexed to \(R\), bracketing the completion-specific claims by which those representations extend or interpret it beyond what is fixed in \(R\).
\end{enumerate}

The third condition does not require that the relevant structure be reducible to directly observable quantities, nor that it admit a formal representation independent of every particular completion. It requires only that the invariant structure can be identified as a determinate content in its own right once completion-specific claims are bracketed, and that its empirical warrant does not depend on choosing one otherwise underdetermined completion. Depending on the case, this may include conservation constraints, symmetry relations, invariant measures, transition structures, or probability and correlation structures over observable histories. In the classical examples considered above, Liouville-measure preservation and its finite-resolution consequences provide instances of such structure; in the quantum case, preparation--intervention--outcome relations supply the corresponding empirical anchor.

By contrast, features associated with the localisation of modal selection---whether in fine-grained initial conditions, stochastic outcomes, collapse mechanisms, or other completion-specific structure---do not acquire model-invariant ontological warrant where empirically admissible formulations differ on these points. Their status is therefore representational relative to \(R\), without prejudging whether or not they are unreal. They may remain scientifically well-motivated hypotheses about how invariant structure extends beyond the established domain, and further evidence may subsequently discriminate between them.

Importantly, the criterion is both fallibilist and regime-relative.\footnote{This regime-relativity concerns the warrant for ontological commitment, not the reality to which that commitment is directed.} It does not deny that presently variant structure may become empirically accessible at finer resolution, over longer timescales, or under different experimental interventions; nor does it preclude preferring particular completions on grounds of simplicity, continuity, explanatory power or integration with wider theory. It distinguishes, however, between corroborated structure within an empirically established regime and extrapolation of that structure into a regime in which competing continuations remain empirically undiscriminated. Where new observations corroborate additional structure, the admissible model class contracts and the ontological commitments licensed by the criterion may correspondingly expand, while earlier commitments may be recontextualised within the wider structure. Misidentification of invariant structure within an empirically well-explored regime likewise remains possible.

Understood in this way, determinism and indeterminism need not mark an invariant metaphysical divide in nature. Where the distinction varies across models belonging to the same empirically admissible class, it reflects alternative ways of representing or completing the same corroborated structure. The task of ontology, while that underdetermination persists, is therefore not to select one such completion by fiat, but to identify and articulate the structure that remains invariant across them.

\subsection{Quantum foundations}

As a roadtest of the model-invariance criterion, I would like to sketch its application to quantum foundations. None of the standard realist approaches is licensed by the model-invariance criterion alone. These interpretations reproduce the empirically established quantum structure only by supplementing it with ontological commitments that vary radically between otherwise empirically equivalent\footnote{Here and below, ``empirically equivalent'' should be understood relative to a specified corroborated regime: the claim is that these formulations reproduce the same empirically established quantum structure within that domain, not that they must remain exactly equivalent under every possible extension or refinement of it. Collapse theories such as GRW may, in principle, yield departures from standard quantum predictions in regimes not yet empirically discriminated.} formulations, while operational or epistemic approaches avoid such commitments without identifying the invariant empirical structure itself as ontological.

Thus Bohmian and other primitive-ontology approaches supplement the wavefunction with definite particle configurations or other local beables and typically deterministic trajectories (Bohm 1952; Allori et al.\ 2008); collapse theories add genuinely stochastic state reduction together with a corresponding primitive ontology (Ghirardi, Rimini and Weber 1986); and Everettian approaches retain universal unitary dynamics while treating branching structure as physically significant (Wallace 2012). These are serious and highly developed realist proposals, but the structures that distinguish them---particle trajectories, stochastic collapses, flashes or mass densities, branching worlds---are precisely those on which empirically equivalent quantum formulations disagree. They therefore cannot, by the model-invariance criterion alone, inherit the ontological warrant attaching to the quantum structure common to those formulations. Conversely, broadly operational or epistemic approaches decline some or all of these additional commitments, but thereby stop short of the positive realist claim made here: that the empirically corroborated invariant structure itself warrants ontological commitment.

Two less standard approaches come considerably closer to the present proposal while making almost opposite representational choices. The thermal interpretation (Neumaier 2019a,b,c) places deterministic expectation-value structure at the centre of its ontology, whereas Barandes' stochastic reconstruction (2025) takes stochastic microstate transitions as fundamental. The contrast between these approaches therefore provides a particularly useful test of whether either modal commitment belongs to the model-invariant content of quantum theory.

The thermal interpretation proposes a fully deterministic
ontology, according to which expectation values of physical quantities evolve
deterministically and constitute the primary ontic structure. On this view
there is no fundamental collapse and no irreducible randomness; apparent
stochasticity arises only at the level of coarse-grained description, when the
detailed dynamics of interactions---typically involving degrees of freedom that
have become effectively inaccessible through decoherence and environmental
coupling in measurement-like contexts---are no longer empirically resolvable. By contrast, Barandes' stochastic--quantum correspondence represents quantum systems in terms of definite configurations undergoing intrinsically stochastic, generally indivisible dynamics. In both approaches, Hilbert-space objects, including the wave function, are secondary representational devices rather than fundamental ontology, and by construction, both reproduce the
standard quantum-statistical predictions. This raises the now familiar question of whether the determinism or
indeterminism of these formulations---together with other allegedly ontic
features such as expectation values, memory, collapse, or trajectory
structure---should be regarded as features of the world itself, or as
representational choices internal to a given formulation.

Barandes' stochastic--quantum correspondence begins from what he calls an indivisible stochastic process. Such a process is specified by a configuration space \(C\) and a family of transition maps \(\Gamma_{t\leftarrow t_0}\), whose entries
\[
\Gamma_{ij}(t\leftarrow t_0)=p(i,t\mid j,t_0)
\]
give the first-order conditional probabilities for transitions between configurations. The point is that an indivisible process does not specify a unique full non-Markovian stochastic dynamics. Rather, it represents an equivalence class of non-Markovian processes sharing these first-order conditional probabilities while potentially differing in their higher-order conditional probabilities and hence in the probabilities assigned to complete trajectories. Barandes argues that the first-order structure suffices to reproduce the empirical predictions of quantum theory, leaving the higher-order and whole-trajectory probabilities unspecified.

He then constructs a Hilbert-space representation of this stochastic structure. A transition matrix \(\Gamma(t\leftarrow 0)\) can be encoded by a complex time-evolution matrix \(\Theta(t\leftarrow 0)\) satisfying
\[
\Gamma_{ij}(t\leftarrow 0)=|\Theta_{ij}(t\leftarrow 0)|^2.
\]
The resulting matrix admits a Kraus decomposition, and by enlarging the configuration space with an ancillary space one can invoke Stinespring dilation to obtain a unitary representation of the enlarged dynamics. In the unistochastic case the correspondence takes the particularly simple form
\[
\Gamma_{ij}(t\leftarrow 0)=|U_{ij}(t\leftarrow 0)|^2,
\]
while the original stochastic transition structure is recovered from the enlarged unitary dynamics by marginalisation over the ancillary configurations. If \(U(t\leftarrow0)\) is differentiable, its self-adjoint generator defines a Hamiltonian and yields the usual Schrödinger and von Neumann equations. On Barandes' reading, however, the Hilbert-space objects thereby introduced are representational devices for the underlying stochastic dynamics rather than its fundamental ontology: the system itself always occupies a definite configuration, while the wave function has a nomological rather than directly ontic status.

From the present perspective, the most suggestive feature of this construction is not simply that it supplies an indeterministic alternative to a fundamentally unitary ontology. Barandes' own formalism already distinguishes between a transition structure sufficient to reproduce the empirical content of quantum theory and additional higher-order or whole-history structure that remains unspecified. In this respect it comes unusually close to the model-invariance strategy developed here. The difference is that Barandes goes on to interpret definite configurations and intrinsically stochastic evolution as fundamental: the Hilbert-space description is treated as a representation of that underlying stochastic ontology. Model invariance instead asks whether the correspondence itself gives us reason to stop one step earlier. If the empirically corroborated transition structure admits both the stochastic representation and its unitary Hilbert-space representation, while the further allocation of modal information differs between them, then that common structure may possess stronger ontological warrant than either representation-specific account of what fundamentally realises it.

Neumaier's thermal interpretation takes a markedly different route. Rather than reconstructing quantum theory from an underlying configuration-space dynamics, it assigns direct physical significance to what he calls q-expectations and q-correlations---the formal expectation values and correlation functions of quantum theory, interpreted not merely as statistical summaries of measurement outcomes but as objective properties of individual systems, whether or not they are directly measurable or empirically accessible. The state of the universe determines the complete hierarchy of such quantities and evolves deterministically according to the quantum dynamics; stochastic behaviour arises only in the description of subsystems whose interaction with their environment cannot be fully controlled (Neumaier 2019c).

This brings the thermal interpretation close to the present proposal in one important respect. It shifts ontological emphasis away from microscopic trajectories and individual measurement outcomes towards expectation-value and correlation structure, which is precisely the kind of relational structure that may remain stable across different empirically accessible levels of description. It also gives a natural account of why macroscopic quantities become increasingly sharp: suitable q-expectations averaged over sufficiently large spacetime regions may have very small uncertainties and thereby support an approximately classical description.

From the present perspective, Neumaier's construction can therefore be understood in a way closely analogous to the role assigned above to a smooth Hamiltonian completion of classical dynamics. In the classical case, dynamical relations that are well corroborated across an accessible range of scales may be smoothly continued under arbitrary refinement, yielding a particularly natural deterministic completion without thereby establishing that this continuation remains exact at indefinitely finer scales. Neumaier's interpretation may be read similarly: the expectation-value and correlation structure of quantum theory, together with its deterministic unitary dynamics, is extended to a universal description in which the complete hierarchy of expectation values and correlations is assigned objective physical significance. This provides a coherent and highly natural deterministic completion of empirically corroborated quantum structure, and may legitimately guide theoretical development beyond the presently accessible domain.

The model-invariance criterion need not reject such a completion. It requires only that the ontological status of its additional structure remain open to revision where that structure has not itself been empirically fixed. Just as a smooth Hamiltonian continuation may correctly represent model-invariant classical structure within its established domain while eventually requiring modification under further refinement, the thermal interpretation may capture genuine invariant quantum structure without the universal ontological status of its full deterministic completion thereby being guaranteed. The issue is therefore not whether Neumaier's representation is legitimate or explanatorily powerful, but which parts of it derive their warrant from structure invariant across empirically admissible formulations and which amount to a particularly natural continuation beyond that structure.

The contrast with Barandes is therefore especially instructive. Neumaier takes deterministically evolving expectation-value and correlation structure as primary, formulating this in a more general coherent-space framework, whereas Barandes takes stochastic transition structure over definite configurations as primary. In both approaches, the Hilbert-space representation is secondary. Both approaches identify substantial structure that plausibly belongs to the model-invariant content of quantum theory, but they place radically different ontological weight on the representations through which that structure is realised. The model-invariance criterion recommends withholding that further commitment unless the difference between these completions becomes empirically or otherwise independently warranted.

From this perspective, the traditional formulation of the measurement problem can also be seen to depend on a prior ontological commitment. The familiar tension between continuous unitary evolution and definite, apparently stochastic outcomes becomes a fundamental inconsistency only if the quantum state and its unitary dynamics are first taken to provide a complete and universally valid description of physical reality. The model-invariance criterion does not show that such a commitment is false, but it asks what independently warrants it when empirically adequate formulations can reproduce the same preparation--intervention--outcome structure while assigning very different ontological significance to unitary evolution, stochastic transitions and measurement outcomes.

As we have seen, however, quantum measurement cannot be treated simply as classical coarse-graining. An observational intervention is itself generally dynamical: a measurement specifies not only probabilities for possible outcomes but also, through the associated instrument, a transformation of the state. The resulting probabilistic transition structure therefore cannot in general be reduced to an epistemic update concerning an independently evolving fine-grained state. At the same time, that operational structure does not by itself determine how the transition is to be completed ontologically. A fundamental collapse process, universal unitary dynamics, or an underlying stochastic or hidden-variable dynamics may each supply such a completion, but none follows from the empirically fixed preparation--intervention--outcome structure alone.

The measurement problem is therefore not so much eliminated as epistemically reclassified. If a universally ontic unitary state is independently warranted, the familiar problem of explaining definite outcomes remains. But the empirical success of quantum theory does not by itself require that starting point. The prior question is which elements of the formalism remain invariant across empirically adequate descriptions of preparations, interventions and outcomes, and which belong only to particular ontological completions. On this view, neither fundamental collapse nor a hidden deterministic or stochastic substrate is demanded by the invariant empirical structure alone. 

\section{Discussion}

It belongs to the nature of empirical science that observed regularities are established relative to idealised regimes in which a limited set of variables is controlled or resolved while other degrees of freedom are neglected. This is not a defect in the relation between theory and experimental inquiry, but reflects the intrinsic nature of empirical warrant itself. Making that regime-dependence explicit in the mathematical formulation of a theory has an important philosophical consequence: it allows us to accord realist weight to structure that is robustly corroborated within a specified domain without requiring that the same structure remain exact under every extension or refinement of that domain. At the same time, alternative completions of the corroborated structure provide mathematically articulated possibilities for investigating how it persists, changes, or breaks down when the empirical regime is extended.

The resulting position is empirically led and fallibilist, but it does not require the success of a theory to be undermined merely because its domain of applicability proves limited. A theory may correctly capture model-invariant features of reality within an idealised regime even if some of its structures require modification elsewhere. Nor need we decide in advance whether every successful theory is ultimately an effective theory or whether some future theory will deserve to be called fundamental. The methodological point is that success within one empirical domain does not by itself warrant extrapolating every feature of a successful model beyond it.

This distinction is particularly important where empirically adequate formulations contain representational structure that is not itself fixed by the evidence. Physics already provides familiar cases in which symmetry-related representatives are treated as alternative descriptions of common invariant structure. The proposal advanced here extends the same methodological principle to equivalence classes defined relative to a specified empirical regime, even where the competing formulations are not related by an exact symmetry and where further empirical refinement may eventually discriminate between them. The resulting freedom is therefore gauge-like: it marks the distinction, at the present empirical level, between structure whose warrant is invariant across admissible completions and additional structure that belongs to particular deterministic, stochastic or hybrid extensions. When new observations discriminate among those extensions, the equivalence class narrows and the warranted ontology may correspondingly expand.

The question ``why?'' is not univocal in the philosophy of physics. In a nomological sense, we may ask why a particular event or regularity follows from a specified dynamical model and its auxiliary conditions. In a structural sense, we may ask why certain relations remain robust across alternative representations or observational regimes. And in a metaphysical sense, we may ask why reality instantiates structures of this kind at all. The present argument concerns principally the second question. A deterministic completion may explain the evolution of a trajectory conditional on its state, while a stochastic completion may specify probabilities over alternative continuations; neither fact by itself establishes that the deterministic or stochastic character of that completion has ontological warrant beyond the regimes in which it has been empirically corroborated.

Where deterministic and indeterministic completions preserve the same empirically corroborated modal structure, neither is thereby fixed as fundamental. Further empirical or theoretical considerations may favour one completion or eventually break the present equivalence class, but while the underdetermination persists, primary ontological commitment attaches to the structure common to the admissible representations rather than to any particular localisation of trajectory-selecting information. It remains possible that there is a fundamental fact of the matter as to whether the underlying dynamics is deterministic or indeterministic. The stronger conclusion of this paper is that the availability of alternative deterministic and stochastic representations makes it reasonable to remain agnostic not only about which representation is fundamental, but about whether determinism and indeterminism \emph{per se} identify a fundamental physical distinction at all.

This conclusion has possible implications beyond the physical cases considered here, though these would require independent development. For example, attempts to infer exhaustive determination of human action from scientific regularities face an analogous danger of moving from empirically established constraints to stronger ontological claims. The present analysis does not extend to addressing the question of agentic freedom, but it helps to identify a more general methodological error at issue in such debates: the passage from structures warranted within an empirical regime to an implicit claim that one particular representation of them exhausts the ontology.

\section*{Acknowledgements}

I would like to thank Nicholas Parkin and Arnold Neumaier for various relevant conversations, and an anonymous reviewer for comments on the original draft of this paper.

\end{document}